\documentclass{kluwer}
\newdisplay{guess}{Conjecture}
\begin{document}
\begin{article}
\begin{opening}
\tolerance=10000 \hbadness=10000
\def\verbfont{\small\tt} 
\def\note #1]{{\bf #1]}} 
\def\be{\begin{equation}}
\def\ee{\end{equation}} 
\def\bearr{\begin{eqnarray}} 
\def\eearr{\end{eqnarray}}
\def\barr{\begin{array}}
\title{ANALYSIS OF HYSTERESIS EFFECT IN {\it p} MODE FREQUENCY SHIFTS AND 
SOLAR ACTIVITY INDICES}
\author{S. C. \surname{Tripathy}}
\author{Brajesh \surname{Kumar}}
\author{Kiran \surname{Jain}}
\author{Arvind \surname{Bhatnagar}}
\runningauthor{S. C. Tripathy et al.}
\runningtitle{HYSTERESIS BETWEEN FREQUENCY SHIFTS AND SOLAR ACTIVITIES}
\institute{Udaipur Solar Observatory, Physical Research Laboratory, Off Bari Road, 
Dewali, P.  B.  No.  198, Udaipur 313001, India 
%(E-mails: sushant@uso.ernet.in, brajesh@uso.ernet.in, kiran@uso.ernet.in, arvind@uso.ernet.in 
} 
\date{}
\begin{abstract}
Using intermediate degree {\it p} mode frequency data sets for 
solar cycle 22, we find that the frequency shifts and magnetic indices show 
a ``hysteresis'' phenomenon. It is observed that the magnetic indices follow 
different paths for the ascending and descending phases of the solar cycle, the 
descending path always seems to follow a higher track than the ascending one. 
However, for the radiative indices,  
the paths 
cross each other indicating phase reversal.
\end{abstract}

\end{opening}

\section{Introduction} 

The solar cycle changes in the Sun's interior are also reflected as a variation 
in the {\it p} mode frequencies. Woodard {\em et al.} (1991)   
showed that the mode frequencies varied on monthly time scales and were correlated 
with the average magnetic flux density on the Sun in short time intervals. 
Bachmann and Brown (1993) pointed out that the frequency shifts were correlated 
differently with magnetic and radiative indices and later confirmed by 
Bhatnagar, Jain and Tripathy (1999).  
A study of solar activity indices by Donnelly (1991) has shown that certain pairs of 
indices follow different paths for the ascending and descending phases of 
the solar cycle displaying a ``hysteresis'' like phenomena. In some cases, these 
observed hysteresis patterns 
start to repeat over more than one solar cycle, giving evidence that this is a 
normal feature of solar activity. Considering seven indices 
of solar activity, Bachmann and White (1994) have shown that hysteresis is present 
among these indices during solar cycle 21 and 22. They further inferred that the 
hysteresis can be approximated as a hierarchy of delay times between the pair 
of indices. This effect has also been investigated by \"{O}zg\"{u}\c{c} and Ata\c{c} (1999) using smoothed 
time series of solar flare index and long duration flare index. 
In cosmic ray intensities in different energy ranges, the hysteresis effect
is also a common feature of cosmic ray modulation (Dorman {\em et al.}, 1999).

Since a pair of activity indices shows the hysteresis effect and the {\it p} mode eigen frequencies 
are correlated with the activity indices, it is expected that the frequency shifts may also 
exhibit a similar hystersis pattern with the activity indices. Some evidence for this 
effect has been recently put forward by the analysis of low degree modes 
(Jim\'{e}nez-Reyes {\em et al.}, 1998), which indicates that the correlation 
between the mode frequencies and magnetic indices varies between the rising 
and falling phases of the solar activity cycle. In similar context, preliminary results for the 
intermediate degree modes have been discussed by Tripathy {\em et al.} (2000). In this paper, we 
carry out a detailed analysis of the intermediate degree modes to observe the characteristics of the 
hysteresis loops. We report, for the first time, that the intermediate degree {\it p} mode frequency shifts 
show a hysteresis like phenomenon with the magnetic activity indices. Thus, we confirm that the 
hysteresis effect prevails not only between the various activity indices but also between the 
activity indices and the global oscillation modes of the Sun.

\begin{table} \caption[]{Frequency data sets}
\label{tab1}
\begin{tabular}[]{|l|l|c|}
\hline 
%\noalign{\smallskip} 
Data source &  Period & No. of Data Sets\\
%\noalign{\smallskip} 
\hline
FTACH/HAO & 17 May 1986 - 10 Nov 1990 & 18\\
BBSO & Mar 1986 - Sep 1990 & 4\\
LOWL/HAO & 26 Feb 1994 - 25 Feb 1996 & 2\\
GONG  & 12 June 1995 - 1 Aug 1999 & 14\\
%\noalign{\smallskip}
\hline
\end{tabular} 
\end{table} 
\section{Observational Data Sets}
Intermediate degree {\it p} mode frequencies have been obtained from the Fourier Tachometer
(FTACH) instrument of 
the High Altitude Observatory (HAO), the Big Bear Solar Observatory (BBSO), the LOWL instrument 
operated by HAO and
the Global Oscillation Network Group (GONG) Project.  
These data sets covering the solar cycle 22 and part of cycle 23,  between 
March 1986 to August 1999,  are summarised in Table~I.
The table lists the 
data sets according to the source, epoch of observation and number of data sets. 
It may be noted that we have used the non-overlapping data sets from the GONG network.

We use four different activity indicators representing magnetic and 
radiative indices  to study the hysteresis effect. 
These  are:  the line-of-sight photospheric magnetic flux  measured from Kitt Peak (KPMI); 
the magnetic plage strength index (MPSI) between 10 and 100 Gauss obtained from Mount Wilson magnetograms 
(Ulrich, 1991);  the equivalent width of HeI 10830 \AA\/ 
line  which primarily originates from the 
chromospheric material (HeI)  and is obtained from Kitt Peak (Harvey, 1984); the integrated 
radio flux at 10.7 cm ($F_{10}$) which originates from several chromospheric-coronal sources,
obtained from Solar Geophysical Data ({\it SGD}).  We also include the unsmoothed International sunspot 
number ($R_{I}$) taken from {\it SGD} for completeness and comparison with earlier results. 

\input epsf
\begin{figure}
\begin{center}
\leavevmode
\epsfxsize=3.5in\epsfbox{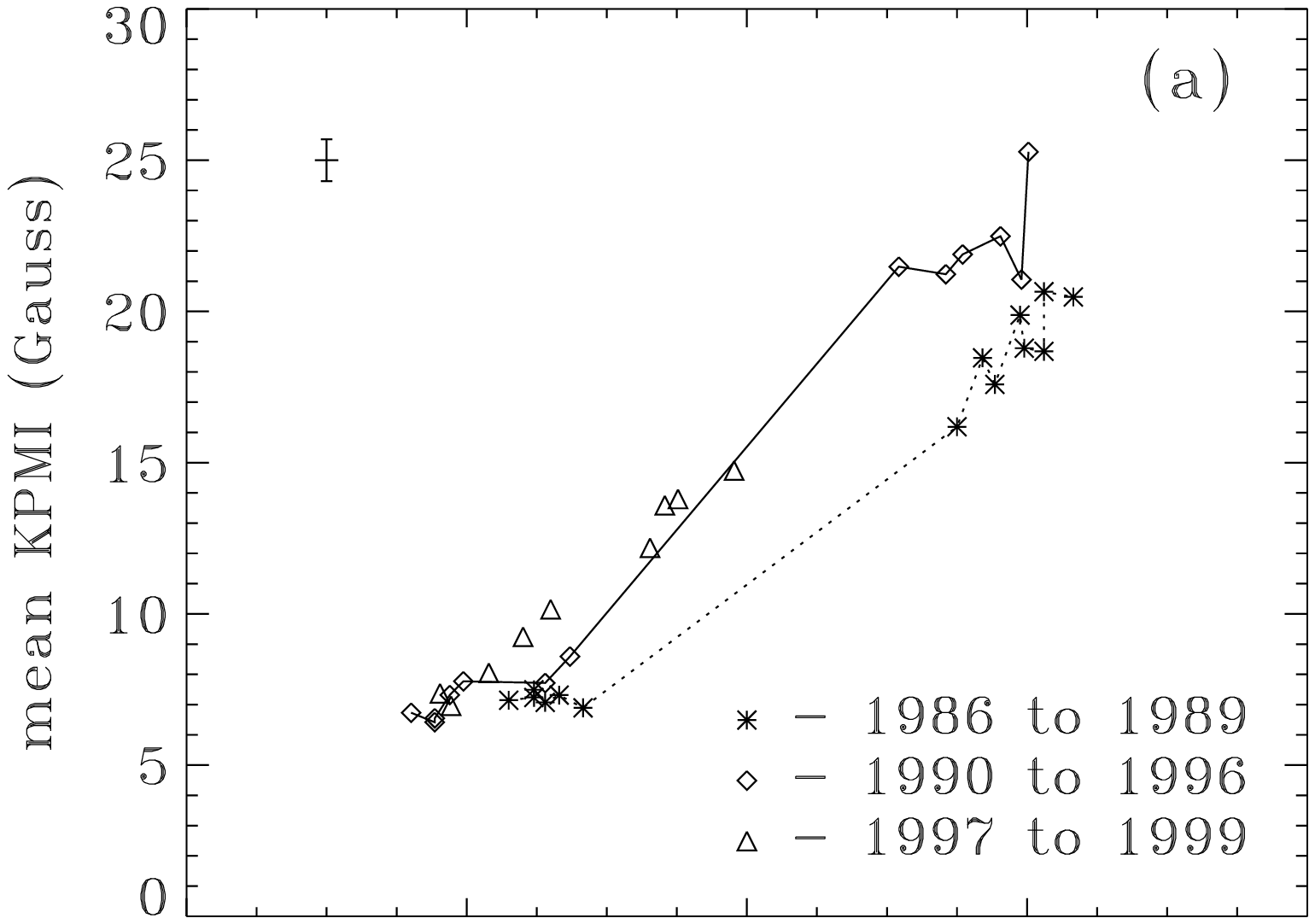}\\
\leavevmode
\epsfxsize=3.5in \epsfbox{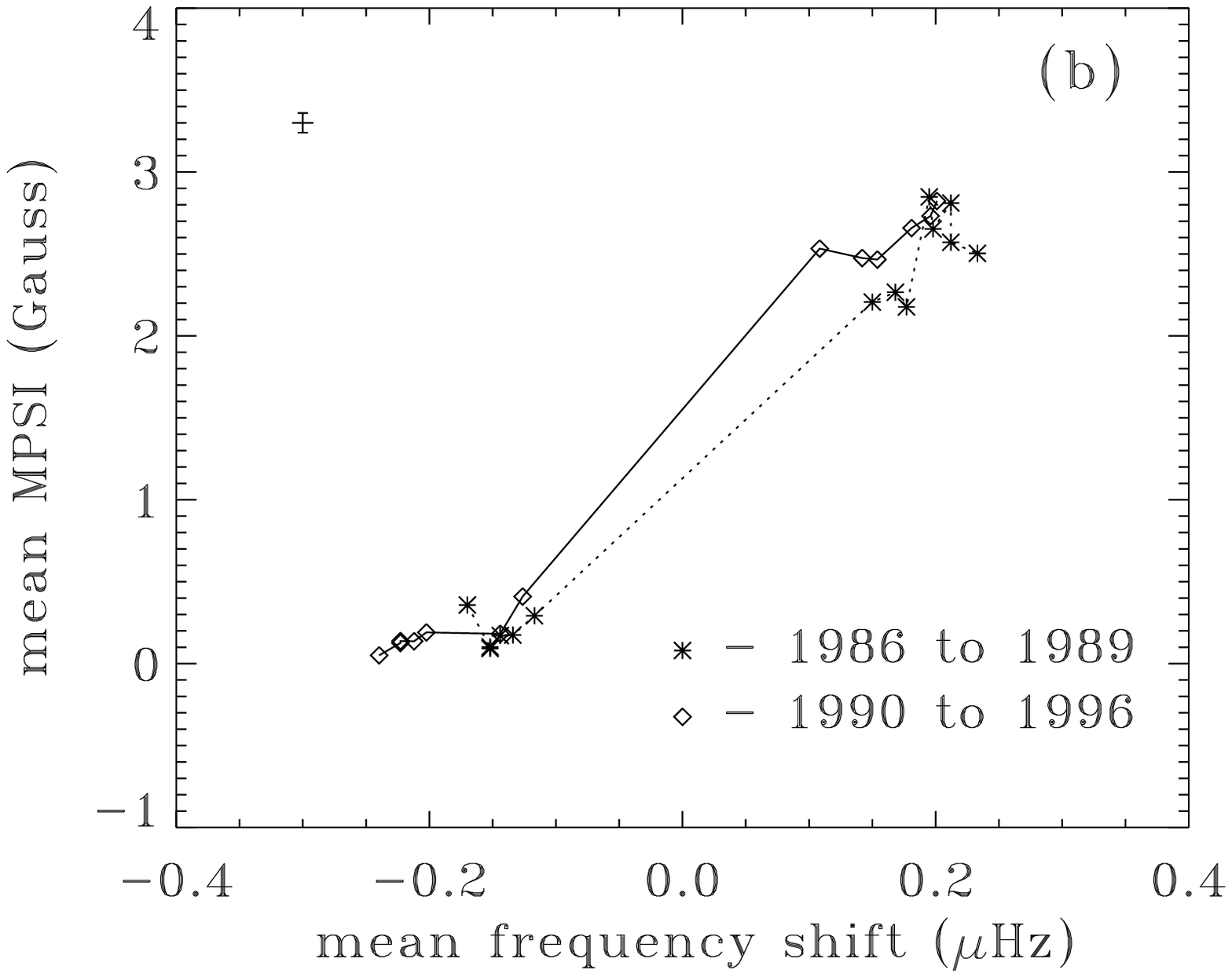}
\caption{Variation of {\it (a)} Kitt Peak Magnetic Index and {\it (b)} Magnetic Plage 
Strength Index with frequency shift. 
It is observed for solar cycle 22 that the descending phase follows a higher 
track than the 
ascending one showing the hysteresis effect. The error-bars at the top left 
corner indicate 1$\sigma$ values.} 
\end{center}
\end{figure}

\begin{figure}
\begin{center}
\input epsf
\leavevmode
\epsfxsize=3.5in \epsfbox{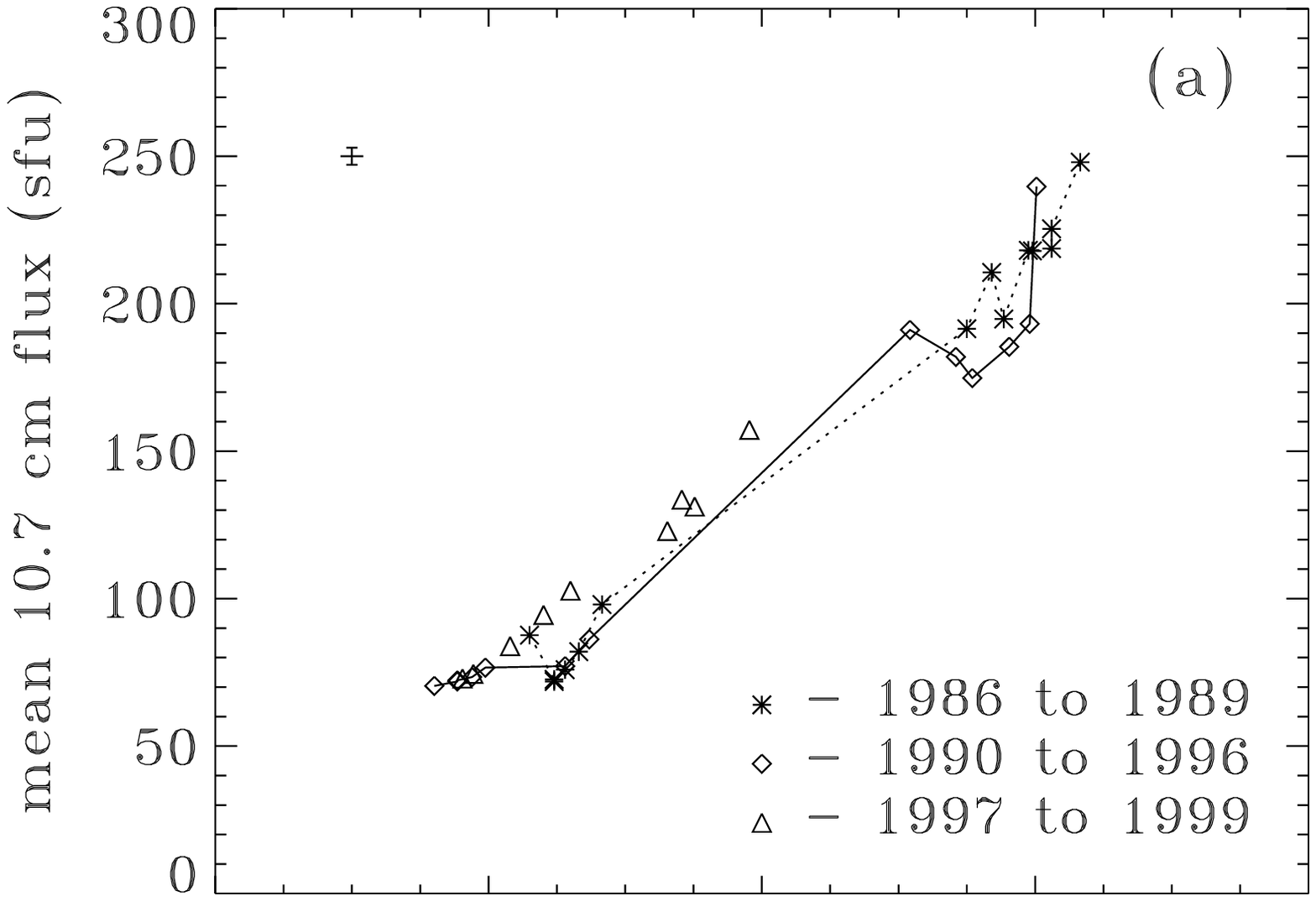}\\
\leavevmode 
\epsfxsize=3.5in \epsfbox{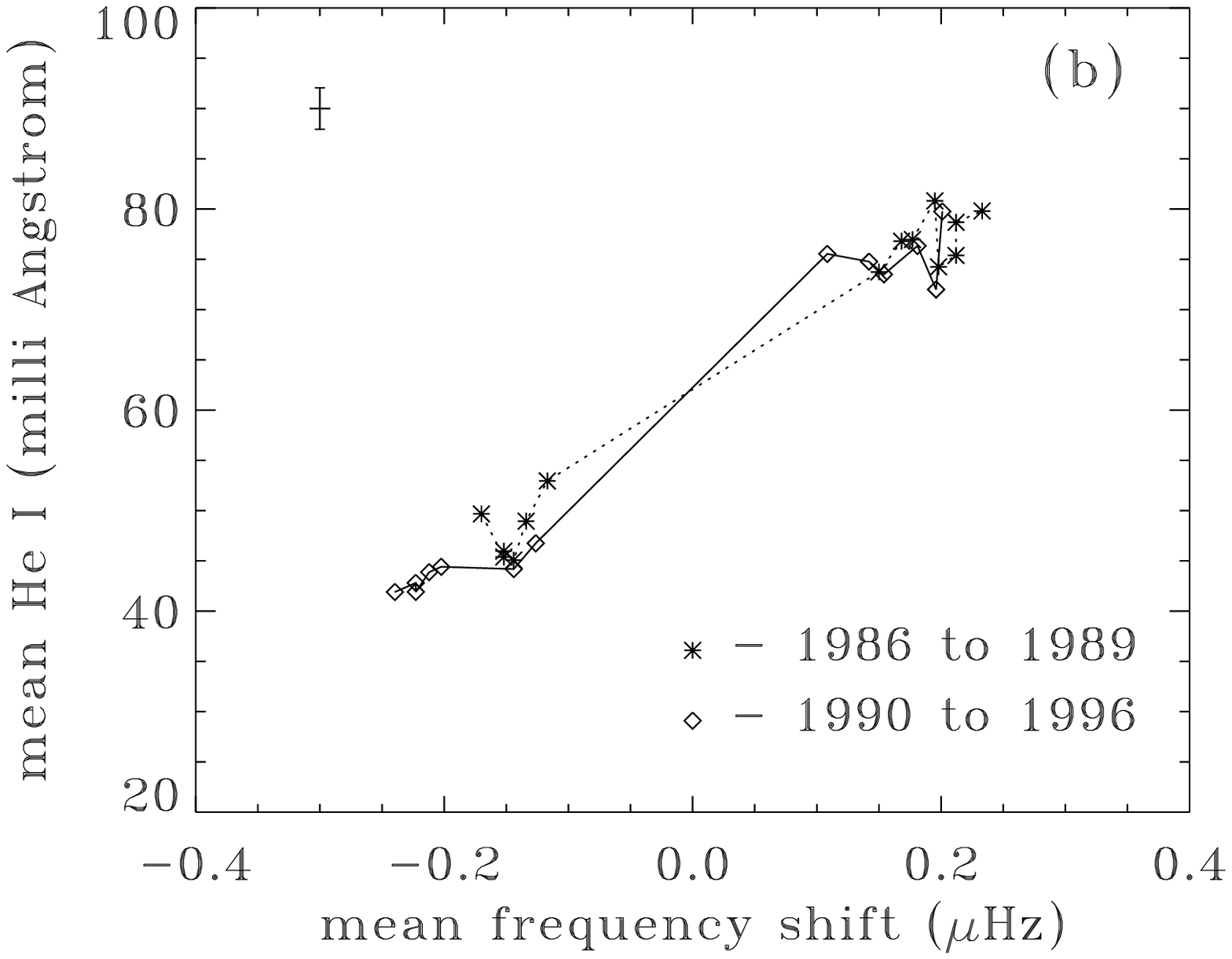} 
\caption{Variation of  {\it (a)} 
10.7 cm radio flux, {\it (b)} Equivalent width of He 10830 \AA\/ line. 
For these indices, the 
 descending and ascending paths cross each other indicating phase reversal. 
The error-bars at the top left corner indicate 1$\sigma$ values} 
\end{center}
\end{figure} 

\section{Results and Discussion}
The centroid frequency shifts are calculated by using the BBSO frequencies 
of 1988 as reference. 
This analysis is restricted to the spherical harmonic degree 
range 20 $\le$ $\ell$ $\le$ 60 and 
frequency range 2600~$\mu$Hz $\le$ $\nu$ $\le$ 3200 $\mu$Hz, due to the direct 
use of frequency shifts from FTACH data as given in Bachmann and Brown (1993).
A mean value, 
together with associated error, is computed for each solar index over the same intervals 
corresponding to the individual frequency shifts measurements, so simultaneous 
values for solar indices and mean frequency shifts are obtained.

Figure~1 shows the variation in mean magnetic field values represented by KPMI and 
MPSI as a function of frequency shifts.  For solar cycle 22,  
both the indices display a hysteresis pattern by following different paths 
 for the ascending and descending phases.  It is also observed that the 
descending path follows a higher track than the ascending one, which is a normal
characteristic of hysteresis in magnetic materials.
 Figure~2 shows similar plots for radiative indices
represented by $F_{10}$  and He I. Here, the ascending and descending paths do not exhibit 
hysteresis patterns, instead the paths cross each other and indicate {\it phase reversal}. 
Similarly, $R_I$  does not 
reveal any hysteresis effect and appears to follow the radiative indices. 

The presence or absence of the hysteresis effect in different activity indices is further 
 confirmed by carrying out the Spearman's rank 
correlation analysis between these indices and  mean frequency shifts. The 
correlation coefficients ($r_{s}$) for the ascending,
descending and the complete solar cycle 22 are summarised in Table~II. In all the 
cases, $F_{10}$ has the maximum correlation while KPMI has the minimum correlation. 
It is evident 
 that the radiative 
indices have a better rank correlation than the magnetic field indices 
represented by KPMI and MPSI while $R_{I}$  is in close agreement with the radiative 
indices. This behaviour of the sunspot number needs to be examined more closely
and is outside the scope of this paper.  

\begin{table} \caption[]{Spearman's rank correlation statistics for solar cycle 22}
\label{tab2}
\begin{tabular}[]{|l|c|c|c|}
\hline 
\noalign{\smallskip} 
Activity & $r_{s}$ & $r_{s}$ & $r_{s}$\\
Index &  (ascending phase) & (descending phase) & (full solar cycle)\\
\noalign{\smallskip} 
\hline
KPMI & 0.87 & 0.91 & 0.79\\
MPSI & 0.83  & 0.96 & 0.88\\
$R_{I}$ & 0.92  & 0.93 & 0.94\\
$F_{10}$ & 0.94 & 0.96 & 0.95\\
HeI & 0.89  & 0.91 & 0.90\\
\noalign{\smallskip}
\hline
\end{tabular} 
\end{table} 
 The trend of variation of frequency shifts with 
activity indices for cycle 23 (upto August 1999) is also shown in Figures~1a and 2a. 
It is apparent that the solar activity level for the ascending phase of  cycle 
23 is higher than cycle 22. However, any conclusion regarding the hysteresis
effect for the current cycle 23 would require consistent data sets for the 
complete cycle e.g., from GONG network.

We have further evaluated the parameter $\oint\Delta\nu$ (Table~III) which represents 
the mean frequency difference between 
the descending and ascending phases of solar activity cycle. The methodology 
for calculating this parameter is the same as that adopted by Jim\'{e}nez-Reyes {\em et al.} (1998). 
In brief, first we omitted the saturation part of the diagrams which are 
obviously at low and high activity. Next, the area enclosed by the closed path 
was calculated and then divided by the value of the common scanned range of 
activity. Thus we obtain $\oint\Delta\nu$ for each diagram together with the 
associated error considering the propagation of the individual errors 
throughout the process. 

\begin{table}
\caption[]{Values of the parameter $\oint\Delta\nu$ for
different  activity indices for cycle~22}
\label{tab3}
\begin{tabular}[]{|l|r|}
\hline 
\noalign{\smallskip} 
Activity Index & $\oint\Delta\nu$\\
&  (nHz)\\
\noalign{\smallskip} 
\hline
KPMI & 110 $\pm$ 8.56\\
MPSI & 60 $\pm$ 8.39\\
$R_{I}$ & 20 $\pm$ 8.36\\
$F_{10}$ & $-$10 $\pm$ 8.93\\
%FI & $-$10 $\pm$ 8.83\\
HeI & $-20$ $\pm$ 8.43\\
\noalign{\smallskip}
\hline
\end{tabular}
\end{table}
We find from Table~III that the value of the parameter $\oint\Delta\nu$ is 
fairly significant for the magnetic indices KPMI and MPSI while for radiative indices
the values are nearly zero. On the other hand,  $R_{I}$ has a small positive value indicating
an intermediate behaviour.

\begin{figure}
\begin{center}
\leavevmode
\input epsf
\epsfxsize=3.5in \epsfbox{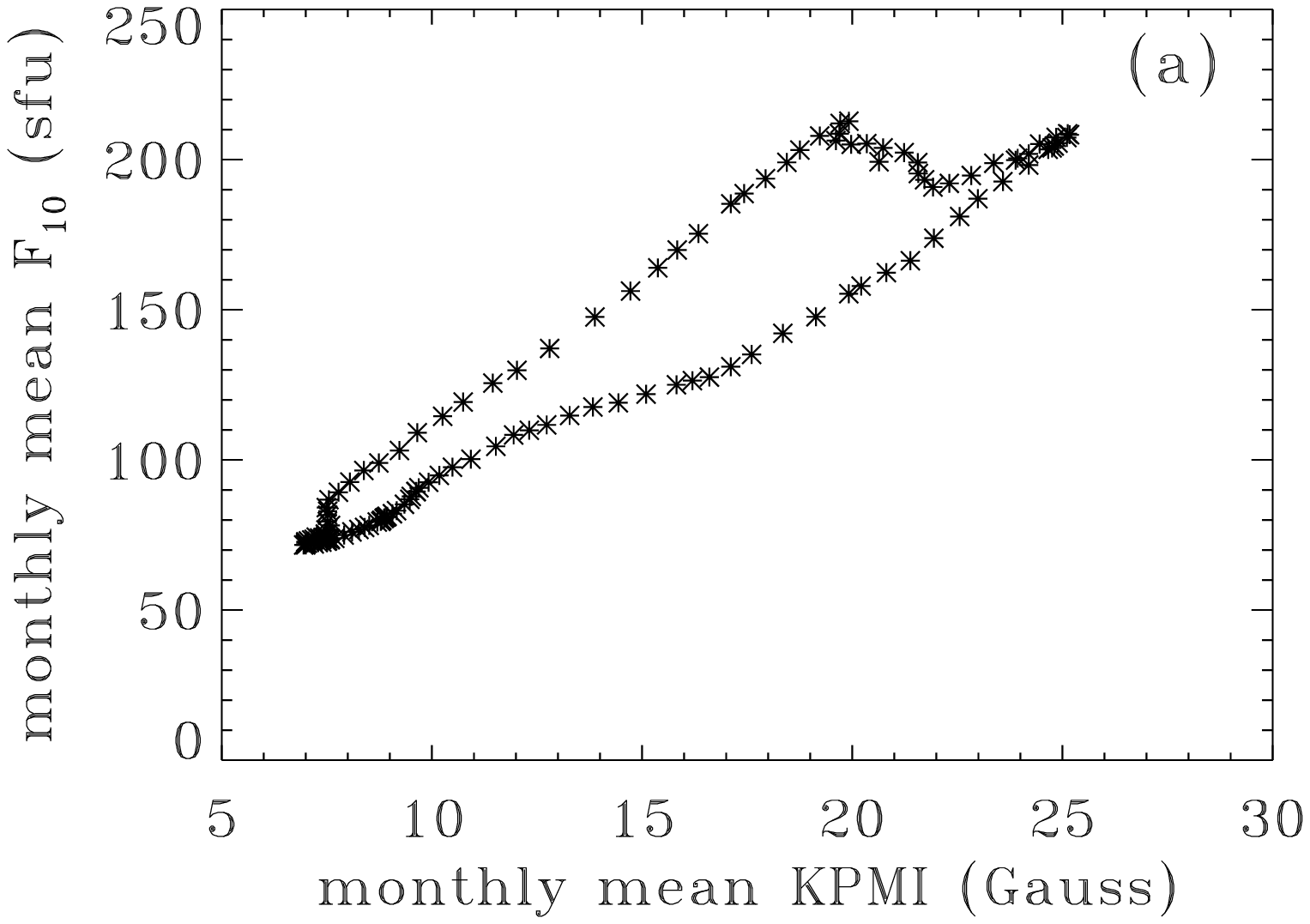}\\
\vspace{0.4in}
\leavevmode
\epsfxsize=3.5in \epsfbox{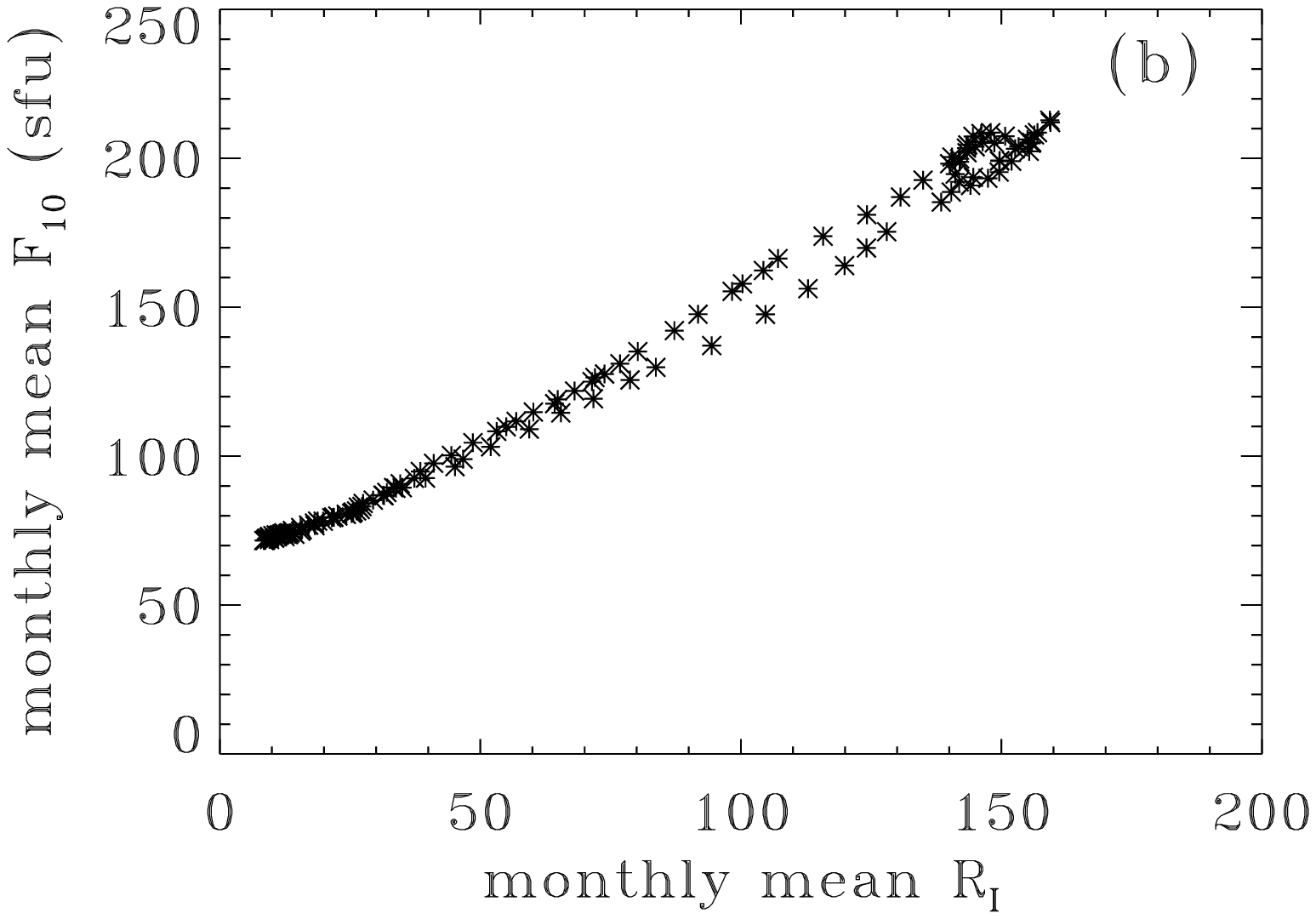}
\caption{Monthly mean plot of {\it (a)} $F_{10}$ versus KPMI and {\it (b)} $F_{10}$ versus 
$R_{I}$ for solar cycle 22. It may be noted that the hysteresis effect is much pronounced 
for the pair of indices $F_{10}$ and KPMI as compared to the pair $F_{10}$ and $R_{I}$. The points 
 in these plots represent the monthly mean of the activity indices on which a running mean of 12 months 
 is applied. } 
\end{center}
\end{figure}

Observationally, it now appears that the {\it p} mode frequency shifts and the activity indices 
exhibit hysteresis phenomenon. A preliminary interpretation for the hysteresis like effect between 
a pair of activity indices has been proposed by Bachmann and White (1994) on the basis of delay times 
behind the leading activity index. This argument is tested by plotting $F_{10}$ as a function of 
both KPMI and $R_I$ (Figure~3) for the solar cycle 22. 
The $F_{10}$ and KPMI show a conspicuous hysteresis loop while this effect is marginally seen between 
$F_{10}$ and $R_I$. This suggests a long time delay between the activity pairs; KPMI -- $F_{10}$ and $R_I$. 
This is supported by the widely accepted fact that the surface magnetic fields preceed in time to the 
appearance of the sunspots.

The hysteresis phenomenon observed in the case of the global oscillation modes can also be interpreted 
as a time delay between the activity indices and the mode frequencies. Thus, the hysteresis effect seen 
between the surface magnetic fields (KPMI and MPSI) and the mode frequencies is attributed to the long 
delay times between them. This is also manifested in the weak correlations between these magnetic indices 
and the {\it p} mode frequency shifts. On the other hand, the insignificant hysteresis effect observed between 
the frequency shifts with the sunspots and the radiative indices denotes small time lag. Thus, we believe 
that the phenomenon of hysteresis may provide an explanation as to why the radiative indices 
have better correlation with the frequency shift than the magnetic 
indices as was earlier pointed out by Bachmann and Brown (1993) and Bhatnagar, Jain and Tripathy 
(1999).

In summary, we find that the intermediate degree frequencies of solar cycle 22 
show a ``hysteresis'' phenomenon with the magnetic field indices whereas no 
such effect is seen in the radiative indices. However, more consistent frequency data sets 
for the current solar cycle would provide 
a better plausible explanation of the hysteresis shapes between the activity indices and 
structural parameters such as {\it p} mode frequencies.

\acknowledgements
The BBSO {\it p} mode data were acquired by K. Libbrecht and M. Woodard, 
Big Bear Solar Observatory, Caltech. LOWL data were obtained from\\
(http://www.hao.ucar.edu/public/research/mlso/LowL/lowl.html),\\
NSO/Kitt Peak magnetic 
and Helium measurements used here are produced 
coperatively by NSF/NOAO; NASA/GSFC and NOAA/SEL. 
This work utilizes data obtained by the Global Oscillation Network
Group (GONG) project, managed by the National Solar Observatory, a
Division of the National Optical Astronomy Observatories, which is
operated by AURA, Inc. under a cooperative agreement with the National
Science Foundation. The data were acquired by instruments operated by
the Big Bear Solar Observatory, High Altitude Observatory, Learmonth Solar 
Observatory, Udaipur Solar Observatory, Instituto de Astrophsico de 
Canaris, and Cerro Tololo Interamerican Observatory. This work is partially 
supported under the CSIR Emeritus Scientist Scheme and Indo-US collaborative 
programme $-$ NSF Grant INT-9710279.

\end{article}
\end{document}